\documentclass[showpacs,preprintnumbers,amsmath,amssymb]{revtex4-1}

\usepackage{graphicx}
\usepackage{dcolumn}
\usepackage{bm}
\usepackage{color}

\begin{document}

\title{Transient and self-limited nanostructures on patterned surfaces}

\author{V. Dimastrodonato}
\affiliation{Tyndall National Institute, University College Cork, ``Lee Maltings,'' Dyke Parade, Cork, Ireland,}
\author{E. Pelucchi}
\affiliation{Tyndall National Institute, University College Cork, ``Lee Maltings,'' Dyke Parade, Cork, Ireland,}
\author{P. A. Zestanakis}
\altaffiliation{Present address: School of Electrical and Computer Engineering, National Technical University of Athens, 15773 Athens, Greece.}
\affiliation{The Blackett Laboratory, Imperial College London, London SW7 2AZ, United Kingdom}
\author{D. D. Vvedensky}
\affiliation{The Blackett Laboratory, Imperial College London, London SW7 2AZ, United Kingdom}


\date{\today}


\begin{abstract}
Site-controlled quantum dots formed during the deposition of (Al)GaAs layers by metalorganic vapor-phase epitaxy on GaAs(111)B substrates patterned with inverted pyramids result in geometric and compositional self-ordering along the vertical axis of the template. We describe a theoretical scheme that reproduces the experimentally-observed time-dependent behavior of this process, including the evolution of the recess and the increase of Ga incorporation along the base of the template to stationary values determined by alloy composition and other growth parameters.  Our work clarifies the interplay between kinetics and geometry for the development of self-ordered nanostructures on patterned surfaces, which is essential for the reliable on-demand design of confined systems for applications to quantum optics.            
\end{abstract} 

\maketitle

\section{Introduction}

Selective epitaxial growth on patterned surfaces has fostered immense development of high-quality ordered semiconductor nanostructures for optoelectronic and integrated quantum optics applications \cite{kapon89,madhukar93,koshiba94,notzel98} due to the direct control over crystal morphology at the nanometer scale. The initial pattern, consisting of adjacent planes with different crystallographic orientations, evolves toward a stationary shape resulting from the interplay between the pattern \emph{and} the growth conditions (growth rate, temperature and material composition), which determine the facet-dependent kinetics \cite{ozdemir92,sato04,heider93}.  This enables the fabrication of morphologically-controlled nanostructures with electro-optical features tuned \emph{remotely} by epitaxial deposition on patterned substrates.

Within this class of nanostructures are pyramidal quantum dots (QDs) and vertical quantum wires (VQWRs), whose optical properties make them exceptionally suitable as optoelectronic devices and quantum information/communication building blocks \cite{kapon04,baier04,mereni09,pelucchi12}. Recent work on arrays of entangled photon emitters \cite{juska} expands the frontier for applications because of the uniformity of the QD field and the intrinsic crystallographic symmetry of individual QDs \cite{dupertuis11}. 

Pyramidal QDs are grown by metalorganic vapor-phase epitaxy (MOVPE) on GaAs(111)B patterned with inverted tetrahedral recesses exposing three (111)A lateral facets \cite{hartmann98}. During growth the morphological evolution of this template is dictated by the relative precursor decomposition, adatom surface diffusion, and interfacet mass transport between the planar (111)B and grooved (111)A surfaces \cite{pelucchi07,pelucchi11,dimas12}. This leads to a variation in the width of the (111)B basal plane and the actual composition of the material. Hence, a new profile forms along the vertical axis of the recess with a {\it self-limiting} width \cite{kapon95} which depends only on the initial pattern, growth conditions and nominal material composition.  

Self-limiting growth in pyramidal recesses, widely observed experimentally \cite{biasiol98,hartmann98}, has been recently modelled \cite{dimas12} with a theory based on phenomenological reaction-diffusion equations for each facet of the initial template. This approach reproduces quantitatively the experimental trends for the variation of the self-limiting profile with growth conditions and alloy (AlGaAs) composition in the stationary regime, e.g.~capillarity-induced broadening of the self-limiting base for higher Ga concentration and growth temperature, along with the segregation of Ga at the base of the template. The accuracy of the model confirms the validity of a growth scenario dominated by the surface kinetics of precursors and adatoms.  However, the \emph{transient} morphology and composition of the epitaxial layers within the pyramidal recess provide a more stringent test of our model because of the evolving interplay between surface kinetics and geometry as the system approaches the steady state.

The most important aspect of the transient morphology is associated with the quantum confinement by the self-limiting profile and the concentration variation along the vertical axis of the inverted pyramid. All spectral properties of pyramidal QDs and VQWRs are the direct result of the morphology produced by the self-limiting mechanism. In order to assure the manipulation, reproducibility and uniformity of these nanostructures, a requisite for practical applications, design criteria (e.g. layer thickness and uniformity, alloy content, etc.) must be advanced. The time-dependent theoretical model we introduce here satisfies these criteria and, moreover, opens up the possibility of engineering of new low-dimensional seeded nanostructures (e.g.~multi-coupled dots, dot-in-a-wire, transient dot shapes), and shows promise for the development of integrated quantum memories and gates and the design of decoherence-proof structures, all of which are strongly influenced by transients.

General time-resolved models of epitaxial layers on planar/patterned substrates have been based on continuum growth equations \cite{kardar86,kan06a,haselwandter10}, canonical rate equations \cite{ozdemir92,keizer87}, three-dimensional continuum theory \cite{sato04,chiu07} and phase-field formulations \cite{eggleston02}. In this paper we present a theory based on the time-dependent solution of the reaction-diffusion equations presented previously \cite{dimas12} to obtain the geometrical and compositional evolution within the pyramidal recess towards the steady state. Atomic force microscopy (AFM) studies, carried out for the first time on samples specifically designed \cite{GC} to analyze transient surface kinetics, confirm the validity of our model for quantitatively describing the variation of the width and segregation effects. 

The outline of this paper is as follows: Sec.~\ref{secII} recalls the fundamental aspects of the reaction-diffusion equations describing the precursors/atom dynamics on patterned substrates \cite{dimas12} and introduces our theoretical approach for the analysis of the time-dependent morphological evolution. In Sec.~\ref{secIII} we present experimental and calculational results which reproduce the time-evolution of both the profile and the accompanying Ga segregation effect. We then discuss the theory and experiments with regard to the geometry of the template and the surface kinetics on the facets of the template in Sec.~\ref{secIV}. Section \ref{secV} presents our conclusions and future directions.

\section{Theory of growth kinetics in a pyramidal recess}  
\label{secII}

\begin{figure}[top]
\includegraphics[width=8.6cm]{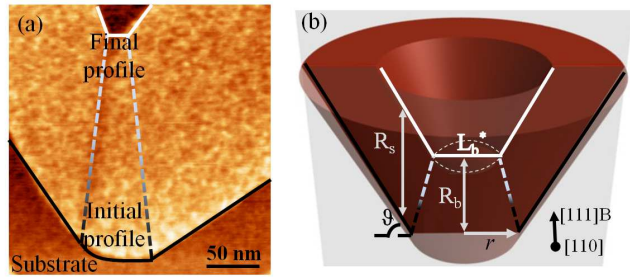}
\caption{(Color online) (a) Flattened \cite{horcas07} cross-sectional AFM scan (height signal) of an Al$_{0.2}$Ga$_{0.8}$As layer (brighter) grown on pure GaAs inside a pyramidal recess. The evolution of the initial (GaAs) profile (black line) towards the self-limited (Al$_{0.2}$Ga$_{0.8}$As) profile (white line) is tracked along the central axis of the recess in a representative way. (b) The pyramidal recess in our conical geometry. The (transparent) cleaving plane crosses the template in the middle, where the self-limited profile $L_b^*$ is indicated. $R_s$ and $R_b$ are the relative growth rates on the sidewalls and bottom, respectively, $\vartheta$ is the angle between (111)B and (111)A, as delivered from the patterning process, and $r$ is the radial coordinate [Eq. (\ref{eq2})].}
\label{fig1}
\end{figure}

Our theoretical approach is based on reaction-diffusion equations for each atomic species migrating on each facet and with matching conditions that account for interfacet mass transfer \cite{dimas12}. The pyramidal template is replaced by a truncated inverted conical recess [Fig.~\ref{fig1}(b)] and MOVPE is modelled by a three-step process: (i) the precursors [trimethylgallium/aluminum(TMGa/Al), for group III, and arsine (AsH$_3$), for group V] reach the substrate surface by diffusion through a boundary layer, where (ii) they diffuse until they decompose at step-edges \cite{shitara93} to release the atomic constituents of the growing material, after which (iii) group-III adatoms diffuse on the growing surface until they are incorporated into the growth front \cite{chua08}.  The surface kinetics that determine the rates of these steps require an accurate parametrization which is able to reproduce the experimentally observed variations of the self-limiting structures with the growth conditions. To simplify the parametrization, we use kinetic parameters\cite{dimas12} that are obtained by fitting the Al concentration-dependence of the self-limiting width in the stationary regime.  Since we deal with AlGaAs/GaAs, no strain contributions are included in our formulation \cite{niu11}.

Figure~\ref{fig1}(a) shows an AFM scan of a representative structure grown in a pyramidal recess \cite{GC1}. The lower and upper darker areas represent the GaAs substrate and GaAs epitaxial layer, respectively, and the brighter thick layer corresponds to the nominal Al$_{0.2}$Ga$_{0.8}$As alloy. By examining the initial and final shape of the template we can track the evolution of the self-limiting profile during growth from a wider to a narrower width.  A schematic illustration of this process is shown in Fig.~\ref{fig1}(b) together with the truncated conical recess geometry we employ for the formulation of the model.

Our theory focusses only on the group-III species (Ga and Al), as none of the kinetics of the group-V species (As) is supposed to be rate limiting. The concentrations ($n_i$) of Ga and Al on each facet (assumed to be independent)  are described by reaction-diffusion equations of the form,
\begin{equation}
\label{eq1}
{\partial n_i\over\partial t}=D_i{\bf \nabla}^2n_i+F_i-\frac{n_i}{\tau_i}\, ,
\end{equation}
with $D_i$ the diffusion constant, $F_i$ the effective atom flux, and $\tau_i$ the adatom lifetime to incorporation on the $i$th facet. For notational simplicity, we have omitted labels for Ga and Al.  These equations are supplemented by matching conditions which impose the continuity of adatom concentrations and fluxes at the boundaries between the sidewalls and bottom facet.

The solution of (\ref{eq1}) enables then us to express the growth rate $R_i$ of each facet  as
\begin{equation}
\label{eq2}
R_i(r,t)=\frac{dz_i}{dt}=\frac{\Omega_0}{\tau_i}n_i(r,t)\, ,
\end{equation}
with $\Omega$ being the atom volume and $r$ the radial coordinate [Fig.~\ref{fig1}(b)]. From the schematic depiction of the self-limiting process in Fig.~\ref{fig1}(b), we derive a differential equation for the evolution of the profile width $L_b$:
\begin{equation}
\label{eq3}
\frac{dL_b}{dt}=2\left(R_b-\frac{R_s}{\cos\vartheta}\right)\cot\vartheta\, ,
\end{equation}
where $R_b$ and $R_s$ are the instantaneous growth rates on the bottom and lateral facets, respectively, calculated from (\ref{eq2}) (note that $R_b$ and $R_s$ are themselves functions of $L_b$ due to the continuity of the boundary conditions).

The direct solution of the time-dependent problem requires the numerical integration of Eqs.~(\ref{eq1})--(\ref{eq3}), but we have adopted an alternative strategy based on incremental {\it stationary} solutions of (\ref{eq1}) over an appropriate time interval. Consider the epitaxial deposition of a single layer and analyze the time scales involved in the adatom kinetics. Equation~(\ref{eq1}) suggests that there are two time scales associated with the diffusion and subsequent incorporation of the atoms. The first is the lifetime to incorporation $\tau_i$ of Ga and Al on each facet, which defines the decay times of their concentrations to the steady state. The second measures the relaxation of the concentrations over the characteristic length of a system. In the specific case of the inverted pyramidal pattern, we identify as characteristic lengths of the system the lengths of the bottom and side facets, yielding the time scales $t_i=L_i^2/D_i$. To ensure that our incremental stationary-based approach is a valid approximation, the incremental time step must be compatible with the foregoing time scales. To this end, we define the dimensionless parameter ${\cal R}_i=L_i^2/D_i\tau_i$ for each facet, as the ratio of the two types of time scale, which can be used to express (\ref{eq1}) in dimensionless form as
\begin{equation}
\label{eq4}
{\partial n_i\over\partial t}={1\over {\cal R}_i}{\bf \nabla}^2n_i+1-n_i\, .
\end{equation}
On the side facet, ${\cal R}_s> 1$ for both species, so the relaxation of the solution to the steady state is determined largely by the adatom lifetimes.  On the bottom facet ${\cal R}_b\gg1$ for Al, but ${\cal R}_b\approx 1$ for Ga, indicating that the time scales for incorporation and diffusional relaxation are comparable.  Hence, by choosing the time step as the maximum of the adatom lifetimes on both facets, we obtain the adatom concentrations from the stationary solutions of (\ref{eq1}) for {\it fixed} $L_b$.  

We have used a time step $\Delta t = 3$~s, with $\tau_{\rm max}\approx 2.4$~s \cite{dimas12} (Table~\ref{table1}).  Smaller time steps, down to $\Delta t=0.5$~s, comparable to the maximum adatom lifetime on the basal facet ($\approx 0.4$~s) also produces acceptable convergence of the solution, demonstrating the dominant role of the kinetics on the basal facet. The stationary solutions to (\ref{eq1}) are then used to calculate the growth rates in (\ref{eq2}) , which produce the incremental change in $L_b$ by integrating (\ref{eq3}), whereupon the cycle is repeated to obtain the evolution of the basal facet.

\section{Evolution of self-limiting profile and alloy segregation}  
\label{secIII}

\begin{figure}[top]
\includegraphics[width=8.6cm]{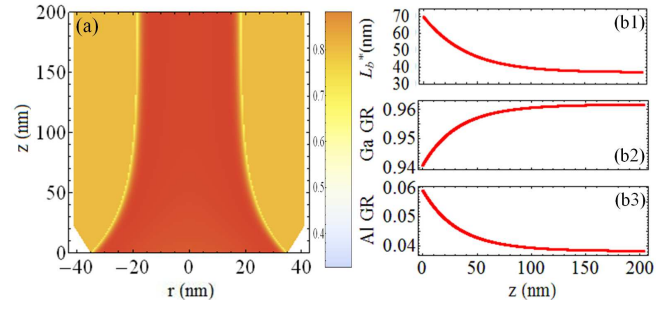}
\caption{(Color online) (a) Density plot of the relative Ga growth rate calculated from (\ref{eq2}) as a function of thickness ($z$) and radial position ($r$) for the structure in Fig.~\ref{fig1}(a). The darker middle branch corresponds to the Ga segregation effect. The color bar indicates the calculated Ga relative growth rate for a nominal Ga content of 0.8.  The discontinuity at the facet boundaries is due to the discontinuity of the kinetic parameters across these boundaries. (b1) Solution of (\ref{eq3}) giving the time (thickness)-dependence of the profile along the axis of the pyramid for the simulated structure in Fig.~\ref{fig1}(a). (b2) and (b3) Ga and Al relative growth rate (GR) along the bottom of the template as a function of the thickness.}
\label{fig2}
\end{figure}

Equations~(\ref{eq1})--(\ref{eq3}) were solved to analyze the dependence of the self-limiting profile width and Ga enrichment on the thickness of the deposited layer. The calculated relative Ga growth rate ($R_i^{Ga}/\left(R_i^{Ga}+R_i^{Al}\right)$) is plotted in Fig.~\ref{fig2}(a), which shows that the trend qualitatively reproduces the approach to the self-limiting profile, together with the strong Ga segregation along the central axis of the pyramidal recess. To better understand the behavior of the self-limiting mechanism along the vertical axis during the transient, we analyze the time-dependence of the self-limiting width and segregation phenomenon at the center of the recess.  Results are shown in Fig.~\ref{fig2}(b). The base of the self-limiting profile [Fig.~\ref{fig2}(b1)] and, in a similar manner, Ga enrichment [Fig.~\ref{fig2}(b2)] and Al depletion [Fig.~\ref{fig2}(b3)] show a significant change over the first hundred of nanometers of the epitaxial layer, before converging toward the steady state at a slower rate. 

\begin{figure}[top]
\includegraphics[width=8.6cm]{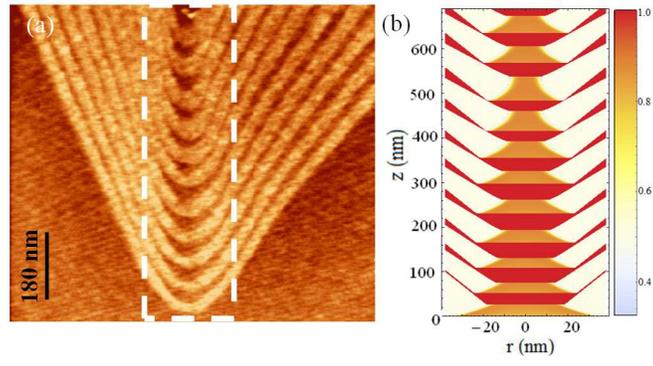}
\caption{(Color online) (a) Height signal of the cross-sectional AFM image of an Al$_{0.5}$Ga$_{0.5}$As/GaAs superlattice structure. GaAs layers appear darker then AlGaAs. The evolution toward the self-limiting profile is tracked along the vertical axis of the pyramid. (b) Density plot of the relative Ga concentration of the simulated Al$_{0.5}$Ga$_{0.5}$As/GaAs superlattice shown in Fig.~\ref{fig3}(a), which reproduces the concentration profile and evolution of the self-limiting width as a function of the layer thickness along the bottom of the template (highlighted area in Fig.~\ref{fig3}(a)).}
\label{fig3}
\end{figure}

\begin{figure}[top]
\includegraphics[width=8.6cm]{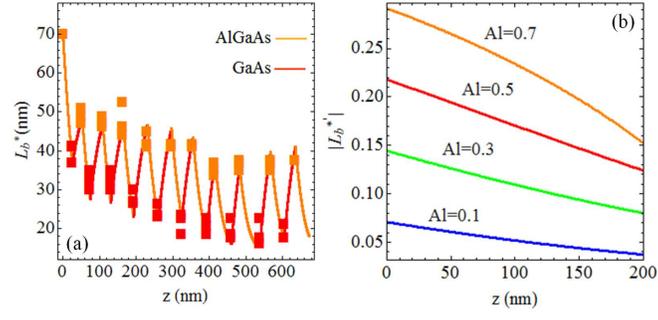}
\caption{(Color online) (a) Calculated solution of (\ref{eq3}) (solid line) for the self-limiting profile as function of the layer thickness for the superlattice structure shown in Fig.~\ref{fig3}(a), and experimental values for initial AlGaAs (orange upper squares) and self-limited GaAs (red lower squares) profile. (b) Rate of the self-limiting width variation $L_b^\ast$ as a function of the layer thickness for different Al concentrations (labeled for each trace).}
\label{fig4}
\end{figure}

To examine the evolution of the self-limiting mechanism in more detail, we designed a specific superlattice structure consisting of a sequence of Al$_{0.5}$Ga$_{0.5}$As layers and GaAs markers \cite{GC2}. A cross-sectional AFM image of this structure is shown in Fig.~\ref{fig3}(a). The insertion of the GaAs markers helps to track the changes of the width of the Al$_{0.5}$Ga$_{0.5}$As layers along the vertical axis of the pyramidal template as thickness increases. We simulated the growth of this structure by solving Eqs.~(\ref{eq1})--(\ref{eq3}) for each layer and imposing appropriate initial conditions: experimental values are used for the thickness of each layer, while the initial profile width is the calculation result of the layer previously deposited. Figure~\ref{fig3}(b) shows the results of the simulation, which reproduce the concentration profile of the layer growth sequence in Fig.~\ref{fig3}(a). For a more quantitative comparison, we measured the initial width for each layer from the AFM scans and plotted the data (squares) in Fig.~\ref{fig4}(a), where we compare with the results of the calculation (solid line).

Apart from demonstrating the agreement between experiment and theory, this comparison adds important information:~(i) the approach to the self-limited profile is a slow process, requiring hundreds of nanometers, depending on growth conditions and material composition [Fig.~\ref{fig4}(a)], before steady values for the profile width and Ga segregation are reached, and (ii) the results corroborate the importance of having a reliable, versatile and efficient theoretical tool to properly conceive nanostructure design and assure reproducible  (lateral) quantum confinement.

\section{Geometry and surface kinetics}  
\label{secIV}

We now consider our calculated results from the standpoint of the interplay between surface kinetics and the geometrical arrangement of the facets that define the pattern on the substrate. In Ref.[\onlinecite{pelucchi11}, \onlinecite{dimas12}] we identified the two main phenomena responsible for the self-limiting profile: the higher lateral growth rate, derived from the faster precursor decomposition rate on the (111)A facets, and the increase of the growth rate on the (111)B basal facet induced by capillarity.  All kinetic parameters are kept constant for the duration of simulated growth. Therefore, the evolution of the profile, together with the development of Ga segregation, results from a changing balance between these two processes. We can therefore explain the transient regime, and the formation of the self-limiting profile, as follows. The decomposition rate anisotropy produces a net flux of adatoms from the sidewalls toward the basal facet of the template where, depending on the composition, adatoms prefer to incorporate (Al) or diffuse (Ga) to balance the adatom concentration gradients relative to the (111)A facets. As the different species rain onto the substrate, diffusion from the sidewalls toward the bottom induces a net flux of adatoms in the opposite direction prior to incorporation along the bottom, which determines a new basal profile. This sequence is repeated, which produces new profile widths as growth proceeds, until the incorporation process completely overcomes the diffusion from the bottom toward the sidewalls. This leads to (i) a constant adatom concentration, (ii) a stationary self-limiting profile and (iii) a higher Ga incorporation rate, since Ga adatoms do not have sufficient time to escape from the bottom facet before incorporation.  Moreover, as Fig.~\ref{fig4}(b) shows, the faster diffusion dynamics and higher incorporation rate for the Al adatoms (Table~\ref{table1}) results in a faster rate of the self-limiting width variation with the thickness as the alloy Al content increases.  Note that the time-dependent analysis allowed us to specifically associate the evolution of the pattern profile evolution with the surface kinetic phenomena, and clearly shows how the latter \emph{adapt} to the changes in the geometry.

\section{Conclusions}  
\label{secV}

In conclusion, we have presented a model which establishes the fundamentals for the understanding and experimental manipulation of the self-limiting mechanism of QD formation within inverted pyramids on patterned surfaces. A key element of our approach is that the parameterization of the model is carried out in the steady state, where the morphology and composition of the QD are in the self-limiting regime.  This parameterization is then used to model the evolution of the QD with a method that side-steps the explicit time-integration of the reaction-diffusion equations. Despite the simplicity of this formulation, our theory reproduces all experimental trends and untangles the growth kinetics on patterned substrates from non-kinetic effects which were believed \cite{biasiolPRL} to be the driving force towards the self-limiting state. Moreover, our work confirms a growth scenario based on the atomistic kinetics of adatom release during precursor decomposition, surface diffusion of adatoms, and their incorporation into the growth front, enabling the design of novel semiconductor nanostructures aimed at integrated quantum information processing.

This research was enabled by the Irish Higher Education Authority Program for Research in Third Level Institutions (2007 to 2011) via the INSPIRE programme, and by Science Foundation Ireland under grants 05/IN.1/I25 and 10/IN.1/I3000. We are grateful to K. Thomas for his support with the MOVPE system.


\clearpage

\begin{table}[t!]
\renewcommand{\arraystretch}{1.2}
\begin{center}
\caption{Adatom lifetime employed in Eq.(\ref{eq1}) at a fixed growth temperature $T_G = 938 K$. Note that $\tau^{-1}$ represents the incorporation rate for each atom.}
\label{table1}
\medskip
\begin{tabular}{cc|r@{.}ll|r@{.}ll}
\hline\hline
Parameter&&\multicolumn{3}{c|}{Al}&\multicolumn{3}{c}{Ga}\\
\hline
$\tau_b^{(k)}$&&0&200&s&0&478& s\\
$\tau_s^{(k)}$&&2&100&s&2&400& s\\
\hline\hline
\end{tabular}
\end{center}
\end{table}

\clearpage

\clearpage

\clearpage

\clearpage

\begin{thebibliography}{99}
\bibitem{GC} All samples were grown in a low-pressure horizontal reactor on substrates patterned by deep UV photolithography and wet chemical etching.
\bibitem{GC1} The structure consists of AlGaAs layers grown on GaAs with Al modulated content. Only the layer with Al concentration of 0.2 is shown here. Growth rate was 0.5~ML/s, V/III ratio $\sim$ 750 and growth (estimated) temperature 938 K.
\bibitem{GC2} The sample was grown with a growth rate of 0.5~ML/s, V/III ratio $\sim$ 750 and growth (estimated) temperature 938 K. The nominal thickness of all layers was modulated from the first to the last loop of growth such to keep an actual constant growth rate along the bottom of the template (as layers approaching the top surface grow faster).
\end{thebibliography}
\end{document}